\documentclass[reprint,amssymb, amsmath, aps, superscriptaddress, showpacs, footinbib,  prb]{revtex4-1}
\pdfoutput=1
\usepackage{graphicx}
\usepackage{color}

\newcommand{\eff}{\text{eff}}
\newcommand{\AFM}{\text{AFM}}
\newcommand{\FM}{\text{FM}}
\newcommand{\eps}{\varepsilon}

\newcommand{\iso}{\text{iso}}

\newcommand{\rr}{\longrightarrow}
\newcommand{\kv}{\mathbf k}
\newcommand{\av}{\mathbf a}
\newcommand{\bv}{\mathbf b}
\newcommand{\cv}{\mathbf c}

\begin{document}

\title{Square-lattice magnetism of diaboleite Pb$_2$Cu(OH)$_4$Cl$_2$}
\author{Alexander A. Tsirlin}
\email{altsirlin@gmail.com}
\affiliation{National Institute of Chemical Physics and Biophysics, 12618 Tallinn, Estonia}
\affiliation{Max Planck Institute for Chemical Physics of Solids, N\"{o}thnitzer
Str. 40, 01187 Dresden, Germany}

\author{Oleg Janson}
\author{Stefan Lebernegg}
\author{Helge Rosner}
\email{Helge.Rosner@cpfs.mpg.de}
\affiliation{Max Planck Institute for Chemical Physics of Solids, N\"{o}thnitzer
Str. 40, 01187 Dresden, Germany}

\begin{abstract}
We report on the quasi-two-dimensional magnetism of the natural mineral diaboleite Pb$_2$Cu(OH)$_4$Cl$_2$ with a tetragonal crystal structure, which is closely related to that of the frustrated spin-$\frac12$ magnet PbVO$_3$. Magnetic susceptibility of diaboleite is well described by a Heisenberg spin model on a diluted square lattice with the nearest-neighbor exchange of $J\simeq 35$~K and about 5\,\% of non-magnetic impurities. The dilution of the spin lattice reflects the formation of Cu vacancies that are tolerated by the crystal structure of diaboleite. The weak coupling between the magnetic planes triggers the long-range antiferromagnetic order below $T_N\simeq 11$~K. No evidence of magnetic frustration is found. We also analyze the signatures of the long-range order in heat-capacity data, and discuss the capability of identifying magnetic transitions with heat-capacity measurements.
\end{abstract}

\pacs{75.30.Et,75.50.Ee,75.10.Jm,91.60.Pn}
\maketitle

\section{Introduction}
\label{sec:intro}
Quantum spin systems show intricate low-temperature phenomena of fundamental\cite{stone2006,giamarchi2008,balents2010,seki2012} and even applied\cite{kagawa2010,wolf2011} interest. Despite strong quantum fluctuations that tend to impair and eventually destroy ordered spin states, most quantum magnets develop long-range magnetic order at sufficiently low temperatures. In two and three dimensions, the lack of the ordered state at zero temperature is only possible in frustrated magnets, where the competition between magnetic couplings amplifies quantum fluctuations. For example, systems of interest are those based on spin-$\frac12$ in the square-lattice or kagom\'e-lattice geometries.\cite{balents2010} The simple square lattice with nearest-neighbor Heisenberg interactions ($J_1$) is non-frustrated. However, a second-neighbor coupling $J_2$ frustrates the system and leads to a rich phase diagram of the $J_1-J_2$ frustrated square lattice model that was extensively studied in the past.\cite{misguich,shannon2004,darradi2008,richter2010,jiang2012}

Experimental search for the $J_1-J_2$ systems on a square lattice remains a challenging problem. While materials with ferromagnetic (FM) $J_1$ can be prepared in a rather systematic fashion using the building blocks of V$^{+4}$ phosphates,\cite{kaul2004,*tsirlin2009,nath2008,tsirlin2011} systems with antiferromagnetic (AFM) $J_1$ are less studied. In the limit of $J_2/J_1\rr 0$, Cu$^{+2}$-based coordination compounds\cite{lancaster2007,*goddard2008,tsyrulin2009,tsyrulin2010,siahatgar2011} and VOMoO$_4$ (Ref.~\onlinecite{carretta2002,*bombardi2005}) are excellent material prototypes of square-lattice systems with weak magnetic frustration. The stronger frustration with $J_2/J_1$ approaching 0.5 has been so far observed only in one material, PbVO$_3$ ($J_2/J_1\simeq 0.35$), that remains controversial because muon spin rotation experiments detect the magnetic ordering below 43~K,\cite{oka2008} but neither thermodynamic measurements nor neutron powder diffraction are capable of observing this magnetic transition.\cite{tsirlin2008} Latest theoretical results suggest that above $J_2/J_1=0.35-0.4$ the $J_1-J_2$ square-lattice system enters the spin-liquid regime, and the magnetic order vanishes.\cite{misguich,darradi2008,richter2010,jiang2012} While PbVO$_3$ may lie on (or be close to) the boundary of this spin-liquid region, further systems with $J_2/J_1\simeq 0.5$ are highly desirable.

The crystal structure of PbVO$_3$ is a tetragonal derivative of the perovskite type. The tetragonal distortion, which is essential for the quasi-two-dimensional (2D) square-lattice magnetism, is only rarely observed in perovskites. Similar structures are found in BiCoO$_3$ (Ref.~\onlinecite{belik2006}) and in the ``supertetragonal'' polymorph of BiFeO$_3$, which is stabilized in thin films.\cite{[{For example: }][{}]bea2009} However, none of these compounds is suitable as a model system for the quantum spin-$\frac12$ square-lattice model because both Co$^{+3}$ and Fe$^{+3}$ bear much higher and largely classical spins. Surprisingly, an appropriate structural analog of PbVO$_3$ can be found in a completely different family of materials. The Cu-based mineral diaboleite, Pb$_2$Cu(OH)$_4$Cl$_2$, has tetragonal crystal structure\cite{rouse1971,cooper1995} that can be derived from the perovskite structure type.

\begin{figure*}
\includegraphics{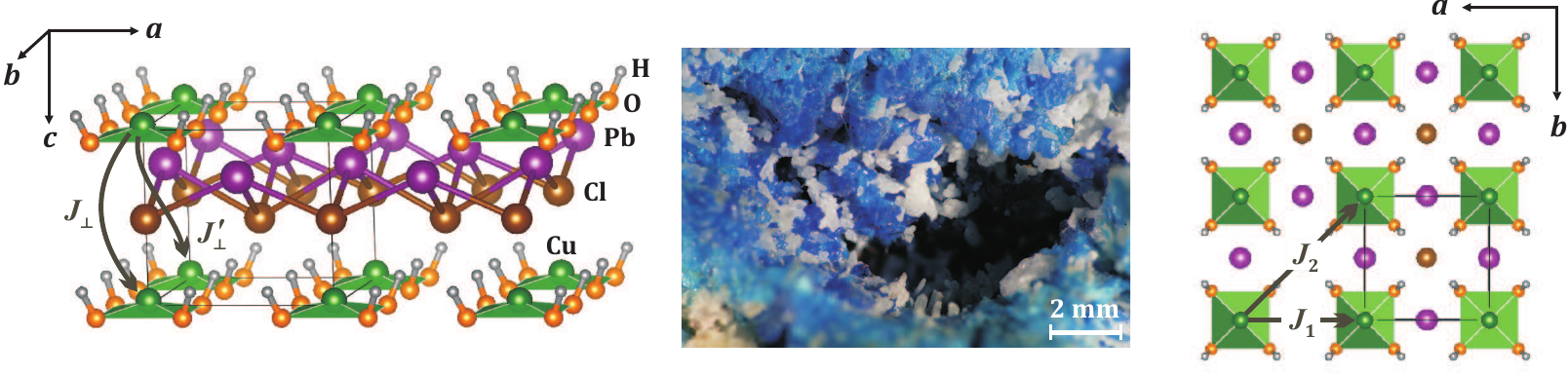}
\caption{\label{fig:structure}
(Color online) An overall view of the diaboleite crystal structure (left panel) and its projection on the $ab$ plane (right panel). Large spheres depict the Pb (violet) and Cl (brown) atoms located between the [Cu(OH)$_4$] magnetic layers. Green spheres inside the plaquettes are Cu atoms. Small spheres show the O (orange) and H (gray) atoms. The middle panel shows the natural sample of diaboleite used in the present study. Note the translucent blue crystals of diaboleite and opaque white crystals of laurionite and phosgenite (see Sec.~\ref{sec:methods} for details).
}
\end{figure*}
The structure of diaboleite features isolated CuO$_4$ plaquettes that form magnetic layers in the $ab$ plane. These layers are interleaved by Pb$_2$Cl$_2$ slabs (Fig.~\ref{fig:structure}). The relation to the parent ABX$_3$ perovskite structure can be understood by writing the chemical composition of diaboleite as Pb$_2$(Cu$\square$)X$_6$, where X = OH, Cl. This way, half of the B-type positions are vacant, with the resulting voids filled by the bulky OH$^-$ anions. While this non-trivial transformation leaves little chemical similarity to perovskites, the crystal structure of diaboleite still features the space group $P4mm$, same as in PbVO$_3$ and PbTiO$_3$. This tetragonal crystal structure is perfectly suited for a quasi-2D magnetism, because the magnetic layers of Cu$^{+2}$ atoms are well separated, with the shortest interlayer distance of 5.5~\r A. The robust tetragonal symmetry entails the perfect square-lattice arrangement of the magnetic sites in the $ab$ plane.

In the following, we present an experimental and computational study of diaboleite. We show that this compound is indeed a good material prototype of the spin-$\frac12$ square lattice. However, its $J_2/J_1$ ratio is very small, hence the magnetic frustration is weak. We discuss the origin of this low $J_2$, and pinpoint another peculiarity, the dilution of the spin lattice through the formation of Cu vacancies that may have implications for experimental studies of the diluted square lattices with spin-$\frac12$.

The outline of this paper is as follows. After summarizing experimental and computational methods in Sec.~\ref{sec:methods}, we report the detailed characterization of the natural mineral sample (Sec.~\ref{sec:sample}) and elaborate on possible deviations from the ideal crystal structure (Sec.~\ref{sec:structure}). We further develop microscopic magnetic models for both the ideal and distorted crystal structures (Sec.~\ref{sec:model}), and apply them to thermodynamic properties of diaboleite (Sec.~\ref{sec:thermo}). Our results are discussed in Sec.~\ref{sec:discussion}, which is followed by a summary and an outlook.

\section{Methods}
\label{sec:methods}
A natural sample of diaboleite (Mammoth-Saint Antony Mine, Pinal Co., Arizona, USA) was provided by the mineralogical collection of the Department of Materials Research and Physics at Salzburg University. A visual inspection of this sample (Fig.~\ref{fig:structure}, middle panel) identified translucent blue crystals of diaboleite mixed with opaque white crystals of other minerals. Both blue and white crystals were detached from the support and separated manually. This way, a 100~mg batch of the diaboleite crystals was obtained.

Phase composition of the samples was determined by laboratory x-ray diffraction (Huber G670 Guinier camera, CuK$_{\alpha1}$ radiation, ImagePlate detector, $2\theta=3-100^{\circ}$ angle range). The batch of the blue crystals mostly contained diaboleite, whereas white crystals were identified as a mixture of phosgenite (Pb$_2$Cl$_2$CO$_3$) and laurionite (PbOHCl). Considering possible contamination of the natural samples with variable -- both crystalline and amorphous -- impurities, sample quality was further checked by bulk chemical analysis and high-resolution x-ray diffraction (XRD). The chemical analysis was performed with the ICP-OES method.\footnote{ICP-OES (inductively coupled plasma optical emission spectrometry) analysis was performed with the VISTA instrument from Varian.} The high-resolution XRD data were collected at the ID31 beamline of the European Synchrotron Radiation Facility (ESRF, Grenoble) at the wavelength of about 0.43~\r A. Details of the experiment are described elsewhere.\cite{tsirlin2011} The \texttt{Jana2006} program was used for the structure refinement.\cite{jana2006} Crystal structures were visualized using the \texttt{VESTA} software.\cite{vesta}

Magnetization measurements were performed with Quantum Design MPMS SQUID magnetometer in the temperature range $2-380$~K in fields up to 5~T. Heat capacity was measured with Quantum Design PPMS in the temperature range $1.8-100$~K in zero field and in the fields of 3~T, 6~T, and 9~T. All measurements were performed on polycrystalline samples.

Magnetic couplings in diaboleite, as well as different aspects of its structural distortions, were analyzed by density-functional theory (DFT) band-structure calculations performed in the \texttt{FPLO}\cite{fplo} and \texttt{VASP}\cite{vasp1,*vasp2} codes that implement basis sets of local orbitals and projected augmented waves,\cite{paw1,*paw2} respectively. Local density approximation (LDA)\cite{pw92} and generalized gradient approximation (GGA)\cite{pbe96} for the exchange-correlation potential were used. Band-structure results are obtained for well-converged $k$ meshes with 336~points in the symmetry-irreducible part of the first Brillouin zone for the crystallographic unit cell, and about 100~points for the supercells. Residual forces in optimized crystal structures were below 0.01~eV/\r A. Effects of strong electronic correlations in the Cu $3d$ shell were treated by introducing the LDA band structure into an effective Hubbard model (model approach), or within the mean-field DFT+$U$ procedure, with the on-site Coulomb repulsion parameter $U_d=6.5$~eV and the Hund's exchange parameter $J_d=1$~eV applying the around-mean-field double-counting correction scheme.\cite{[{For example: }][{}]lebernegg2011,*tsirlin2012} Details of the computational procedure are described in Sec.~\ref{sec:model}.

Thermodynamic properties and N\'eel temperatures of the spin models relevant to diaboleite were obtained from quantum Monte-Carlo (QMC) simulations based on the \texttt{loop}\cite{loop} and \verb|dirloop_sse|\cite{dirloop} algorithms of the \texttt{ALPS}\cite{alps} simulation package. Simulations were performed on 2D ($L\times L$) and 3D ($L\times L\times L/5$) finite lattices with $L\leq 80$ and $L\leq 50$, respectively. Size convergence for the thermodynamic properties (magnetization and magnetic specific heat) was carefully checked. The magnetic ordering temperatures ($T_N$) were determined by calculating temperature dependence of the spin stiffness $\rho_s$ for different $L$ and applying the appropriate scaling procedure, as previously described in Refs.~\onlinecite{sengupta2009} and~\onlinecite{tsirlin2011b,*tsirlin2012b}.

\section{Results}

\subsection{Sample characterization}
\label{sec:sample}
The high-resolution XRD pattern of the natural diaboleite sample reveals strong reflections of the tetragonal diaboleite phase along with few very weak reflections that are attributed to the impurities of phosgenite (Pb$_2$Cl$_2$CO$_3$) and wherryite [Pb$_7$Cu$_2$(SO$_4)_4$(SiO$_4)_2$(OH)$_2$]. The amounts of these impurities are 1.1(1)~wt.\% and 0.4(1)~wt.\%, respectively, according to the Rietveld refinement.\cite{supplement} While phosgenite is a diamagnet, no information on the magnetism of wherryite is presently available. Nevertheless, trace amounts of this impurity should not affect any of the results presented below.\footnote{Edge-sharing CuO$_4$ plaquettes in the crystal structure of wherryite suggest similarities to linarite [PbCu(OH)$_2$SO$_4$], Li$_2$CuO$_2$, and other $J_1-J_2$ chain antiferromagnets: see, e.g., A. U. B. Wolter~\textit{et al.} Phys. Rev. B 85, 014407 (2012).} 

\begin{figure}
\includegraphics{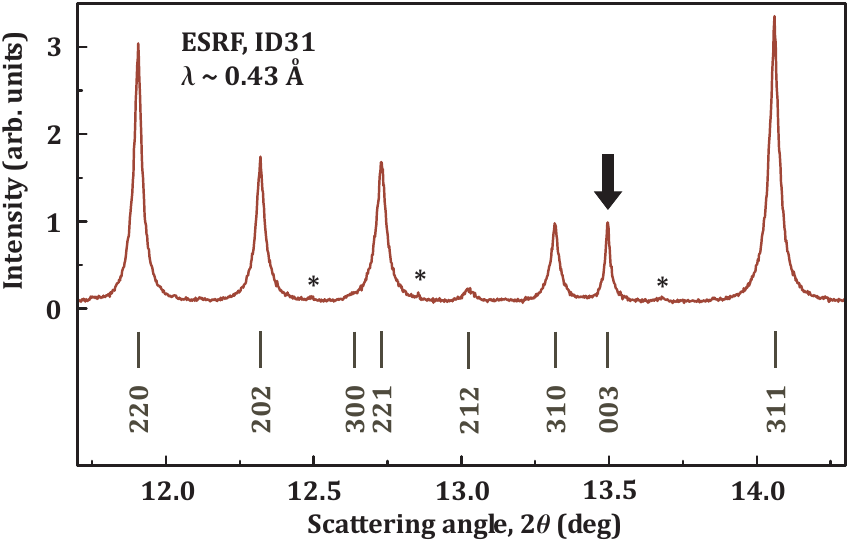}
\caption{\label{fig:xrd}
(Color online) High-resolution XRD pattern of the natural sample of diaboleite. Ticks show the reflection positions, whereas asterisks mark tiny reflections of the phosgenite impurity. The reflections of the second impurity, wherryite, are not visible in this angular range. Note that the 003 reflection is narrower than the reflections with non-zero $h$ and/or~$k$.
}
\end{figure}
Chemical analysis delivered the following bulk composition of the natural sample: 68.1(1)~wt.\% Pb, 9.3(1)~wt.\% Cu, and 0.2(1)~wt.\% Ba. Other elements amenable to the ICE-OES determination (including transition metals, but excluding O, H, and Cl) are below the detection limit of 0.1~wt.\%. These estimates are in reasonable agreement with the expected chemical composition (67.2~wt.\% Pb, 10.3~wt.\% Cu), although slight deviations in the Pb and Cu content, as well as the presence of Ba, signal subtle variations in the stoichiometry of the diaboleite phase. 

The Rietveld refinement of the atomic positions resulted in reasonable atomic displacement parameters (ADPs) of Pb, Cl, and O, $U_{\iso}\simeq 0.01$~\r A$^2$ (the hydrogen position was not refined).\cite{supplement} However, the notably higher ADP of Cu ($U_{\iso}\simeq 0.035$~\r A$^2$) suggested the missing scattering density at the Cu position. Considering the nearly isotropic thermal ellipsoid ($U_a^{\text{Cu}}\simeq 0.03$~\r A$^2$, $U_c^{\text{Cu}}\simeq 0.04$~\r A$^2$ along the $a$ and $c$ directions, respectively), the large ADP of Cu can not be explained by a displacement of Cu from its position on the four-fold axis, and should be rather understood as the formation of vacancies. Indeed, as the ADP was fixed to $U_{\iso}=0.01$~\r A$^2$, the refined Cu occupancy converged to 0.942(4) indicating the deficiency of Cu atoms in the diaboleite structure.\footnote{Note that the ADP and occupancy factor have similar effect on the scattering density and can not be refined simultaneously.} The formation of Cu vacancies conforms to the underestimated Cu content from the chemical analysis that yields about 10~\% Cu deficiency in the diaboleite sample.

The formation of Cu vacancies is justified by both powder XRD and chemical analysis. Indeed, the missing scattering density in XRD could be otherwise explained by the substitution of a lighter element in the Cu position. However, the chemical analysis does not show detectable amounts of any foreign elements that are capable of replacing Cu in the diaboleite structure. Trace amounts of Ba are likely related to the Pb/Ba substitution, which is favored by the similar ionic radii of these elements. The amount of the substituted Ba is well below 1~\%, so it does not show up in the XRD refinement.

Another conspicuous feature of the natural diaboleite sample is the sizable broadening of the $hkl$ reflections with non-zero indices $h$ and/or $k$. This effect is well seen in Fig.~\ref{fig:xrd} where, for example, the 310 reflection is much broader than the neighboring 003 peak.\cite{supplement} Anisotropic reflection broadening can be understood as the formation of stacking faults in the layered crystal structure of diaboleite. This problem is further addressed in Sec.~\ref{sec:structure}, where we also elaborate on the nature of Cu vacancies in the diaboleite structure.

\subsection{Details of the crystal structure}
\label{sec:structure}
The structure refinement puts forward several issues regarding details of the atomic arrangement that may be important for understanding the magnetism of diaboleite: i) positions of hydrogen atoms that are not precisely determined by XRD; ii) formation of Cu vacancies; iii) stacking faults that underlie the reflection broadening.

The position of hydrogen has been determined from single-crystal XRD\cite{cooper1995} using the soft constraint on the O--H bonds (the O--H distance of 0.98~\r A). We further refined the hydrogen position by a DFT-based structure optimization, with all atoms other than hydrogen fixed to their experimental positions. Both LDA and GGA optimizations\footnote{Note that we do not use DFT+$U$ because Hubbard-type electronic correlations introduced in this approach are only relevant to the Cu site, which is kept fixed in the relaxation.} converged to very similar results. The hydrogen atom is in the $4d$ position ($x,x,z$) with $x=0.2975/0.2972$ and $z=0.1191/0.1179$ in LDA/GGA, respectively. The resulting O--H distance is 0.99~\r A, and the Cu--O--H angle is about $112.3^{\circ}$. These parameters only slightly differ from those found experimentally: the O--H distance of 0.98~\r A (constrained) and the Cu--O--H angle of $109.1^{\circ}$.

\begin{figure}
\includegraphics{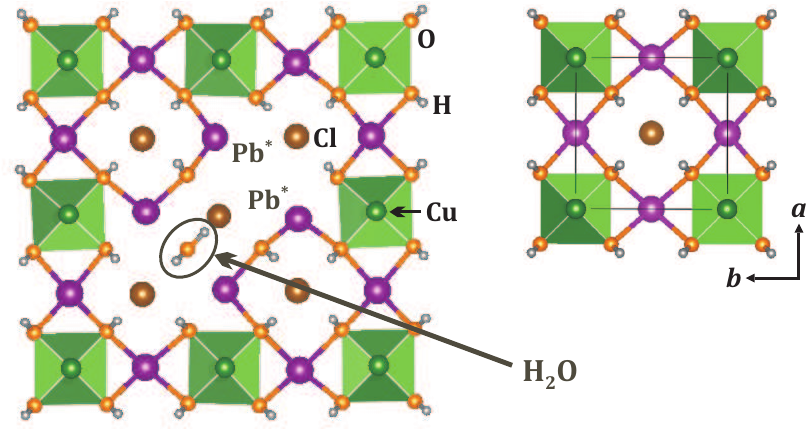}
\caption{\label{fig:vacancy}
(Color online) Left panel: formation of Cu vacancies in the magnetic layer of diaboleite ($\frac14$ of Cu atoms and the respective amount of oxygens are removed). Top right panel: the pristine structure of stoichiometric diaboleite. The removal of one Cu and one O atom releases a water molecule. Additionally, the coordination of Pb atoms changes from four longer Pb--O bonds (2.45~\r A, top right panel) to two shorter Pb$^*$--O bonds (2.32~\r A and 2.40~\r A, respectively; left panel). Only the Pb--O bonds shorter than 2.5~\r A are shown. The notation of atoms follows Fig.~\ref{fig:structure}.
}
\end{figure}
The formation of Cu vacancies can be explained as follows. The removal of one Cu atom creates an uncompensated negative charge that should be balanced by the removal of one oxygen or two chlorines. As the oxygen atoms are strongly bonded to Cu with the Cu--O distance of 1.97~\r A, the removal of oxygen is expected. The remaining hydrogen atom may join one of the OH groups to form a water molecule. Regarding the four oxygen atoms that surround the vacancy, one is removed to compensate the charge, one oxygen forms the water molecule, whereas the two remaining OH groups are still bound to Pb. The resulting structure has been fully relaxed within GGA+$U$ to yield the configuration shown in Fig.~\ref{fig:vacancy} where we consider the $2\times 2$ supercell with one out of four Cu atoms removed. This supercell with 25\,\% Cu deficiency should be representative for the actual diaboleite sample that features $5-10$\,\% of Cu vacancies. The calculations for lower concentrations of vacancies would require even larger supercells, which are hardly feasible for an accurate computational treatment.

The formation of Cu vacancies enhances the \mbox{Pb--O} bonding and reduces the coordination number of Pb atoms around the vacancy. In stoichiometric diaboleite, each Pb atom is surrounded by four oxygens with the Pb--O separations of 2.45~\r A. In Cu-deficient diaboleite, the Pb$^*$ atoms (Fig.~\ref{fig:vacancy}) are strongly bound to only two O atoms, but the respective Pb--O distances are shortened to 2.32~\r A and 2.40~\r A, respectively. The bond to the third oxygen atom extends to 2.53~\r A, whereas the fourth bond is either lost (for Pb$^*$ atoms adjacent to the oxygen vacancy) or stretched up to 2.63~\r A (for the oxygen atom of the water molecule). This way, the lengths of the \mbox{Pb--O} bonds are redistributed, and bonding requirements of Pb can be satisfied, even though one out of four oxygen atoms is removed from the crystal structure. The void formed between the two Pb$^*$ atoms (Fig.~\ref{fig:vacancy}) is typical for Pb$^{+2}$ oxides. It can be ascribed to the ``localization'' of $6s^2$ lone pairs that form a non-bonding region in the crystal structure.\cite{abakumov2011}

The formation of Cu vacancies can be considered as the removal of CuO from the diaboleite structure. The local transformation is written as follows:
\begin{equation*}
  \text{Pb}_2\text{[Cu(OH)}_4\text{]Cl}_2\longrightarrow \text{[Pb}_2\text{(OH)}_2\text{]Cl}_2\text{(H}_2\text{O)}+\text{CuO}.
\end{equation*}
This formal equation reflects the fact that the complex anion [Cu(OH)$_4]^{2-}$, with four OH groups attached to the Cu atom, is transformed into the complex cation [Pb$_2$(OH)$_2]^{2+}$, where the two remaining OH groups are attached to Pb. To evaluate the energetics of Cu vacancies, we use a similar equation and replace the fictitious (and presumably unstable) [Pb$_2$(OH)$_2$]Cl$_2$(H$_2$O) compound with the aforementioned metastable diaboleite having 25\,\% of Cu vacancies (Fig.~\ref{fig:vacancy}). Using fully relaxed atomic configurations of this Cu-deficient diaboleite, pristine stoichiometric diaboleite, and CuO (tenorite),\cite{tenorite1970} we obtain the energy of $+0.27$~eV/f.u. (about 10~kJ/mol) for the formation of 25\,\% of Cu vacancies. Although rather large, this additional energy can be partially tolerated by the entropy term and by chemical potentials of different ions under specific growth conditions. Note that Pb$_2$Cu(OH)$_4$Cl$_2$ is prepared in water solution\cite{winchell1968} and likely follows a similar route of crystal growth in the natural environment. 

\begin{figure}
\includegraphics{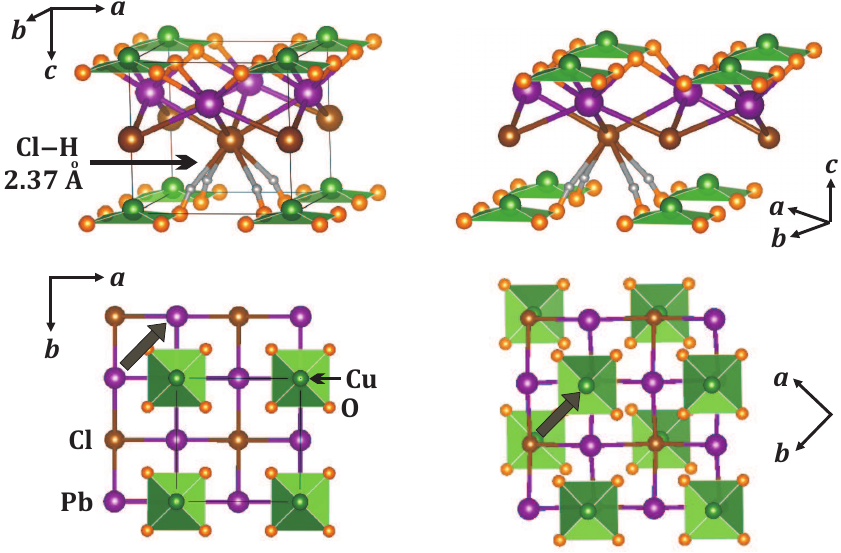}
\caption{\label{fig:stacking}
(Color online) Possible mechanism of stacking faults in the diaboleite structure. The notation of atoms follows Fig.~\ref{fig:structure}. Left panels: parent tetragonal crystal structure. Right panels: optimized structure with the uniform shift of the [Cu(OH)$_4$] layers for $\kv=(\frac12\frac120)$. Note that the [Pb$_2$Cl$_2$] slabs are invariant to this translation, which is denoted by the thick arrow. The weak Cl$\cdots$H hydrogen bonds are the only interaction between the [Pb$_2$Cl$_2$] slab and the neighboring [Cu(OH)$_4$] layer. Therefore, the structures with $\kv=0$ (left panel) and $\kv=(\frac12\frac12 0)$ (right panel) marginally differ in energy. In the bottom panels, hydrogen atoms are omitted for clarity.
}
\end{figure}
Finally, we consider the problem of stacking disorder. The structure of diaboleite is formed by [Pb$_2$Cl$_2$] slabs that are linked to the [Cu(OH)$_4$] units via the Pb--O bonds. On the opposite side of the slab, the connection to the next [Cu(OH)$_4$] magnetic layer is restricted to weak H$\cdots$Cl hydrogen bonds (Fig.~\ref{fig:stacking}, top left panel). Each Cl atom is bonded to four hydrogen atoms of the neighboring unit, with the H--Cl distances as large as 2.37~\r A. This value approaches the upper limit of O--H$\cdots$Cl distances reported in the literature.\cite{aakeroy1999,*steiner2002} The weak interlayer interaction enables regular shifts of the neighboring [Pb$_2$Cl$_2$][Cu(OH)$_4$] layers, and facilitates the stacking disorder. 

In the tetragonal diaboleite structure, the layers are stacked on top of each other (displacement vector \mbox{$\mathbf k=0$}). An alternative stacking sequence could be based on $\kv=(\frac12\frac12 0)$ because the [Pb$_2$Cl$_2$] slabs and even oxygen atoms are invariant to this translation, which only changes the order of Cu atoms and vacancies (note the Pb$_2$Cu$\square$X$_6$ formula discussed in Sec.~\ref{sec:intro}) along with hydrogen atoms. The resulting structure is monoclinic, with the twice larger unit cell and the $Cm$ symmetry (Fig.~\ref{fig:stacking}). In GGA+$U$, it lies only 40~meV/f.u. above the ideal tetragonal structure. This small energy difference is comparable to the entropy term and allows for the formation of stacking faults that are indeed observed experimentally.

To understand the effect of these stacking faults on the reflection broadening, we note that the stacking sequence with  $\kv=(\frac12\frac12 0)$ leads to a monoclinic unit cell with the lattice vectors $\av'=\av+\bv$, $\bv'=\av-\bv$, and $\cv'=\frac{\av+\bv}{2}+\cv$, where $\av$, $\bv$, and $\cv$ are lattice vectors of the parent tetragonal unit cell. The $a'b'$ plane of the monoclinic cell is fixed as the plane of the diaboleite layer. However, the $c'$ axis can have two different orientations, $\frac{\av+\bv}{2}\pm\cv$. Its position also depends on the precise value of the monoclinic angle between $a'$ and $c'$. Therefore, in the reciprocal space the direction of the $(\cv')^*$ axis is fixed [$\,(\cv')^*\perp\av',\bv'$, i.e., $(\cv')^*\|\cv\,$], whereas the direction of the $(\av')^*$ axis is variable. This explains why the $(00l)$ reflections remain narrow, although the reflections with large $h$ and $k$ substantially broaden. Indeed, the refinement of the anisotropic strain broadening yields the largest $hk0$ component, which is related to the variable position of the $(\av')^*$ axis.\cite{supplement}

Although our detailed study of the crystallographic issues is largely motivated by effects observed in the natural sample of diaboleite, similar features are likely relevant to the Pb$_2$Cu(OH)$_4$Cl$_2$ phase in general. The layered nature of the crystal structure facilitates the stacking disorder, whereas the formation of Cu vacancies can be instrumental in tuning low-temperature magnetism of this material. In the following, we report experimental data on thermodynamic properties of diaboleite and analyze them from the microscopic viewpoint by considering the parent tetragonal crystal structure of stoichiometric diaboleite along with possible deviations from this ideal atomic arrangement.

\subsection{Microscopic magnetic model}
\label{sec:model}
\begin{figure}
\includegraphics{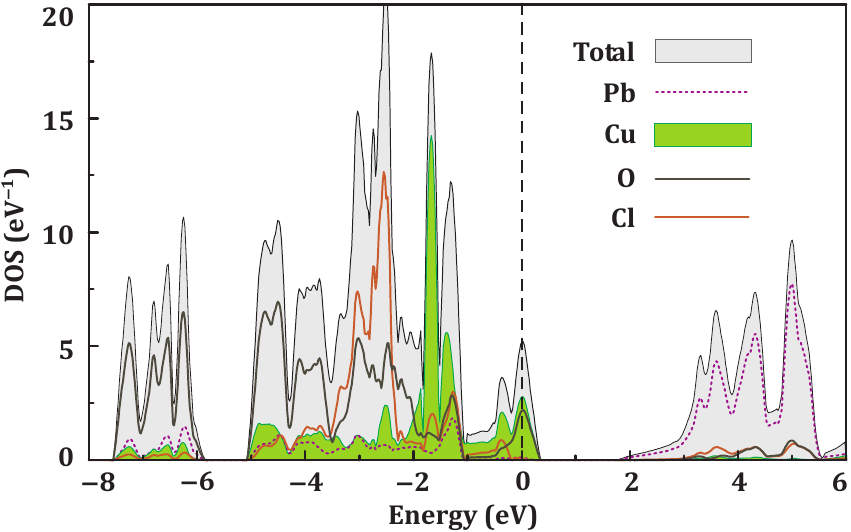}
\caption{\label{fig:dos}
(Color online) LDA density of states (DOS) for the stoichiometric diaboleite. The Fermi level is at zero energy.
}
\end{figure}

The LDA energy spectrum of diaboleite is typical for a cuprate compound. The valence band is dominated by oxygen $2p$ states, with sizable contributions of Cl $3p$ around $-3$~eV, and Cu $3d$ above $-2$~eV (Fig.~\ref{fig:dos}). The large underestimate of strong electronic correlations leads to the spurious metallic energy spectrum in LDA. The LSDA+$U$ and GGA+$U$ calculations yield the band gap of about 2.7~eV in accord with the blue color of diaboleite crystals (Fig.~\ref{fig:structure}, middle panel).

\begin{figure}
\includegraphics{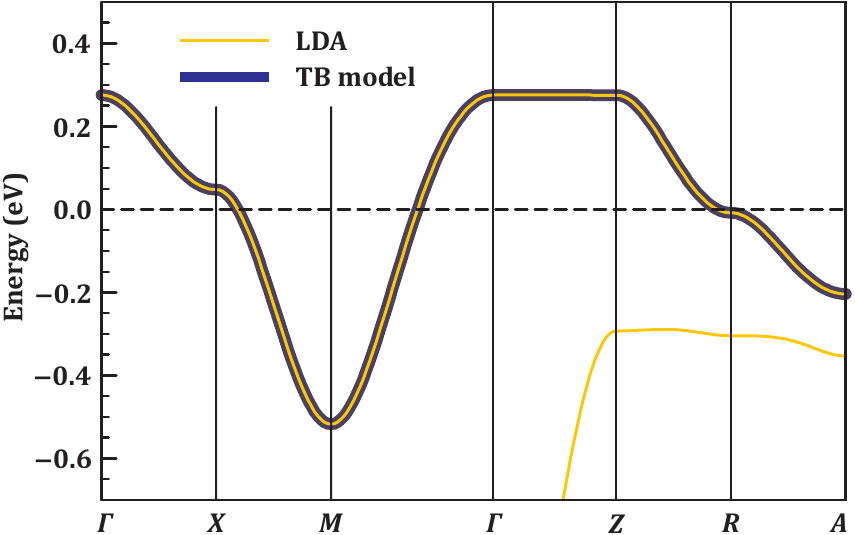}
\caption{\label{fig:band}
(Color online) LDA band structure of stoichiometric tetragonal diaboleite (thin light lines) and the fit with the tight-binding model for the Cu $d_{x^2-y^2}$ band (thick dark line). The $k$ path is defined as follows: $\Gamma(0,0,0)$, $X(0.5,0,0)$, $M(0.5,0.5,0)$, $Z(0,0,0.5)$, $R(0.5,0.5,0.5)$, $A(0.5,0,0.5)$ in units of $2\pi/a$, $2\pi/b$, and $2\pi/c$, respectively.
}
\end{figure}

The Fermi level is crossed by a single band (Fig.~\ref{fig:band}), which is of Cu $d_{x^2-y^2}$ origin according to the conventional crystal-field levels of a strongly elongated octahedron (the in-plane Cu--O distances are 1.97~\r A, whereas the out-of-plane Cu--Cl distances are 2.55~\r A and 2.95~\r A). The Cu $d_{x^2-y^2}$ band is well reproduced by a tight-binding model with the leading nearest-neighbor intraplane hopping $t_1=78$~meV and several long-range interactions that are all weak (Table~\ref{tab:couplings}). Note that the coupling between the magnetic layers is provided by both $t_{\perp}$ (along the $c$ direction) and $t_{\perp}'$ (along [101]). Other hoppings are below 2~meV and can be safely neglected in the minimal microscopic model.\footnote{The sizable hopping $t_{\perp}'$ leads to a deceptive picture in the tight-binding analysis. Normally, one uses the energies at the $\Gamma$, $X$, and $M$ points for a preliminary evaluation of $t_2/t_1$. For $t_2=0$, $\eps_{\Gamma}-\eps_X=\eps_X-\eps_M$, which is not the case for diaboleite (i.e., a sizable $t_2$ should be expected). However, the deviation from the simple $t_1$ square lattice is due to both $t_2$ and $t_{\perp}'$ that are indistinguishable in the $ab$ plane of the reciprocal space ($\Gamma-X-M$). The hopping $t_{\perp}'$ connects each Cu site to 8 neighbors and has strong effect on the dispersion, whereas $t_2$ is small.}

Introducing the LDA hoppings into a Hubbard model with the effective on-site Coulomb repulsion $U_{\eff}=4.5$~eV,\cite{lebernegg2011,tsirlin2012} we identify Pb$_2$Cu(OH)$_4$Cl$_2$ as a magnetic insulator with $t/U_{\eff}<0.02$. Therefore, perturbation theory in $t_i/U_{\eff}$ enables the evaluation of AFM couplings as $J_i^{\AFM}=4t_i^2/U_{\eff}$ to the lowest (second) order. We find $J_1^{\AFM}=63$~K and a very weak frustration by the second-neighbor coupling ($J_2^{\AFM}/J_1^{\AFM}<0.01$). Weak interlayer couplings $J_{\perp}$ and $J_{\perp}'$ render diaboleite a good quasi-2D magnetic system.

The above estimates are based on the model approach that yields the AFM couplings only. However, magnetic couplings -- especially those that are short-range -- may also include a FM component. This FM component can be taken into account by total-energy calculations (supercell approach) where energies of collinear spin states are mapped onto the Heisenberg model to yield full exchange couplings $J_i$ (Table~\ref{tab:couplings}). Then the FM component is evaluated as $J_i^{\FM}=J_i-J_i^{\AFM}$. In the case of diaboleite, this procedure only slightly changes the microscopic scenario. According to Table~\ref{tab:couplings}, $J_1$ remains the leading interaction in diaboleite, whereas other couplings are weak and AFM.

\begin{table}
\caption{\label{tab:couplings}
Cu--Cu distances (in~\r A), transfer integrals $t_i$ (in~meV), and exchange couplings $J_i$ (in~K) in the tetragonal stoichiometric diaboleite. The AFM components are obtained as $J_i^{\AFM}=4t_i^2/U_{\eff}$ with $U_{\eff}=4.5$~eV, whereas the $J_i$ values are calculated using the supercell approach. The superexchange pathways are shown in Figures~\ref{fig:structure} and~\ref{fig:wannier}.
}
\begin{ruledtabular}
\begin{tabular}{lcrcc}
        & Cu--Cu distance & $t_i$ & $J_i^{\AFM}$ & $J_i$  \\
  $J_1$        & 5.88     &  78   &    63        &   38   \\
  $J_2$        & 8.32     & $-7$  &    0.5       &   0.5  \\
  $J_{\perp}$  & 5.50     & $-10$ &    1.0       &   1.0  \\
  $J_{\perp}'$ & 8.05     &   9   &    0.8       &   0.4  \\
\end{tabular}
\end{ruledtabular}
\end{table}

\begin{figure}
\includegraphics{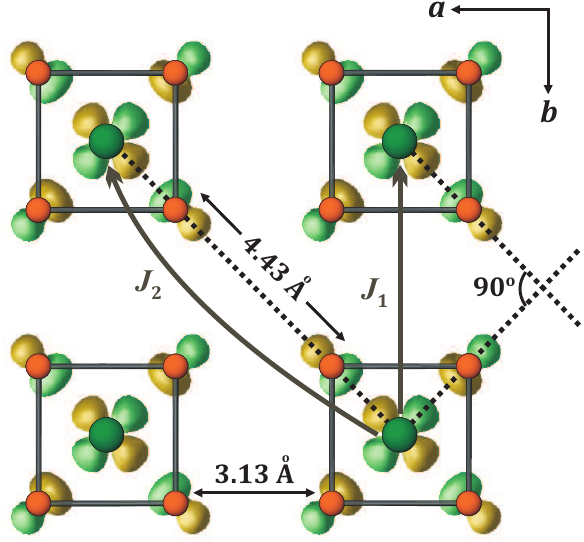}
\caption{\label{fig:wannier}
(Color online) Cu $d_{x^2-y^2}$-based Wannier functions, and the scheme of Cu--O$\cdots$O--Cu superexchange pathways in diaboleite. Different colors show different signs of the Wannier function. Larger (green) and smaller (orange) sheres denote the Cu and O atoms, respectively. The nearest-neighbor coupling $J_1$ features the $90^{\circ}$ orbital overlap and shorter O$\cdots$O distances (3.13~\r A), whereas $J_2$ corresponds to the $180^{\circ}$ overlap and longer O$\cdots$O distance (4.43~\r A).
}
\end{figure}
To elucidate the origin of magnetic couplings in diaboleite, we calculated the Wannier functions (WF) based on the Cu $d_{x^2-y^2}$ orbital character. Each WF (Fig.~\ref{fig:wannier}) includes the Cu $d_{x^2-y^2}$ orbital along with the $\sigma p$ orbitals of oxygen atoms. The superexchange pathway of $J_1$ involves the $90^{\circ}$ overlap of the neighboring WFs (Fig.~\ref{fig:wannier}), which generally leads to a FM contribution to the exchange.\cite{[{See, for example: }][{}]anderson1963,*mazurenko2007} Indeed, we find a sizable FM component $J_1^{\FM}=J_1-J_1^{\AFM}=-25$~K (Table~\ref{tab:couplings}). The second-neighbor coupling $J_2$ features a more favorable $180^{\circ}$ pathway. However, the O--O distance of 4.43~\r A is too large for an efficient overlap of the oxygen orbitals, and the resulting coupling is very weak.

Finally, we evaluate magnetic couplings for the stoichiometric diaboleite with the displaced layers (Fig.~\ref{fig:stacking}) and for the diaboleite with $\frac14$ of Cu vacancies (Fig.~\ref{fig:vacancy}). The formation of vacancies does not disrupt the couplings between the remaining Cu sites: the intralayer hoppings $t_1$ are in the range $74-79$~meV, which is very close to $t_1=78$~meV for the stoichiometric compound. The effect of stacking faults is also weak. However, the change in the stacking sequence should modify the regime of interlayer exchange by replacing $J_{\perp}$ and $J_{\perp}'$ with a single coupling $J_{\perp}''$, which is somewhat weaker than the interlayer couplings in the tetragonal structure (compare $t_{\perp}''\simeq 3$~meV with $t_{\perp}$ and $t_{\perp}'$ of $\pm 10$~meV in Table~\ref{tab:couplings}). The role of interlayer couplings in diaboleite is further discussed in the next section.

\subsection{Thermodynamic properties}
\label{sec:thermo}
\begin{figure}
\includegraphics{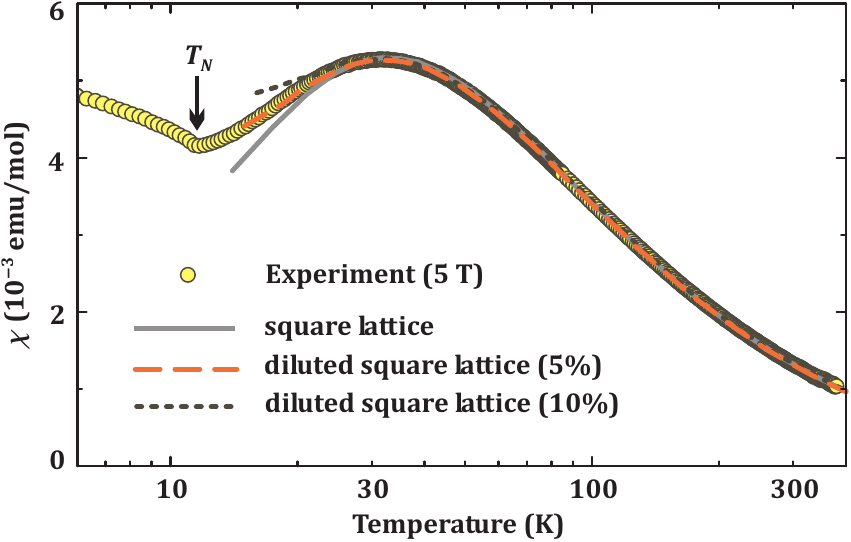}
\caption{\label{fig:chi}
(Color online) Magnetic susceptibility of diaboleite measured in the applied field of 5~T (circles) and QMC fits of three spin models (lines): the stoichiometric square lattice and the square lattices with 5\% and 10\% of vacancies. The arrow denotes the N\'eel temperature $T_N\simeq 11.5$~K at 5~T.
}
\end{figure}

\begin{figure}
\includegraphics{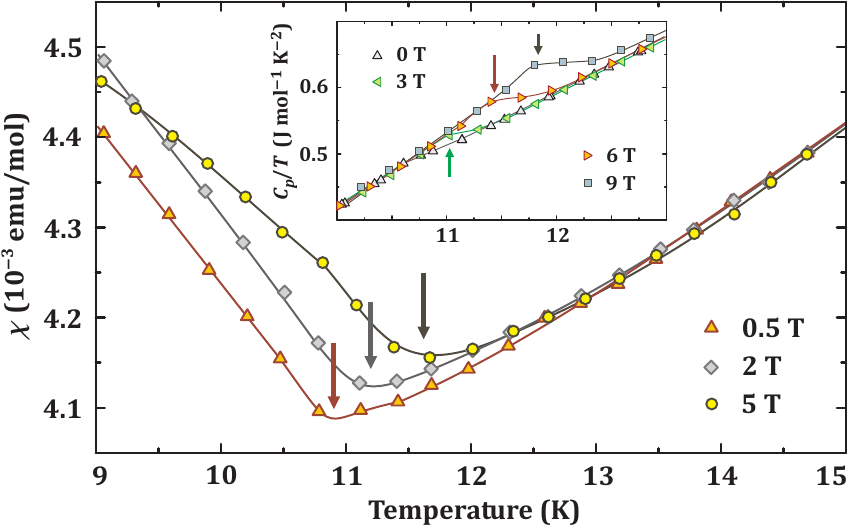}
\caption{\label{fig:chi-field}
(Color online) Field dependence of the N\'eel temperature in diaboleite, as seen from the magnetic susceptibility and specific heat (inset) data. Lines are guide for the eye. Arrows denote the transition temperatures (see text for details).
}
\end{figure}

Magnetic susceptibility of diaboleite shows a maximum at $T^{\max}\simeq 32$~K followed by a kink at $T_N=11-12$~K (Fig.~\ref{fig:chi}). This kink is somewhat field-dependent (Fig.~\ref{fig:chi-field}) and indicates the onset of long-range magnetic order. Above 200~K, the susceptibility follows the Curie-Weiss law
\begin{equation*}
  \chi=\chi_0+\dfrac{C}{T+\theta},
\end{equation*}
with $\chi_0=-2\times 10^{-4}$~emu/mol, $C=0.512$~emu~K/mol, and $\theta=35$~K. The sizable temperature-independent diamagnetic contribution is likely related to the contribution of the sample holder and core diamagnetism. The $C$ value yields the effective magnetic moment $\mu_{\eff}=2.02$~$\mu_B$, which is notably higher than 1.73~$\mu_B$ expected for \mbox{spin-$\frac12$}. The deviation of $\mu_{\eff}$ from the spin-only value implies $g\simeq 2.32$, which is rather high but still in the reasonable range for Cu$^{+2}$ compounds (see, e.g., Ref.~\onlinecite{zvyagin2002,*wolter2012}). The positive $\theta$ value implies predominantly AFM exchange couplings.

The susceptibility maximum can be approximated by a QMC fit for the purely 2D Heisenberg model of the \mbox{spin-$\frac12$} square lattice (solid line in Fig.~\ref{fig:chi}). We find $J=35$~K, $g=2.34$, and $\chi_0=-2\times 10^{-4}$~emu/mol in excellent agreement with the high-temperature Curie-Weiss fit. However, the square-lattice model fails to describe the data below the susceptibility maximum. In this region, the fit can be improved by considering a diluted square lattice with randomly distributed vacancies. The dilution of spin-$\frac12$ sites with vacancies increases the susceptibility below $T^{\max}$ and eventually blurs the maximum. We find the best fit for the lattice with 5~\% of vacancies, $J=37$~K, and nearly unchanged $g$ and $\chi_0$ parameters. This result perfectly matches $J_1=38$~K from DFT (Table~\ref{tab:couplings}). The concentration of vacancies is well in line with the XRD data that yield about 6\,\% Cu deficiency (Sec.~\ref{sec:sample}). 

\begin{figure}
\includegraphics{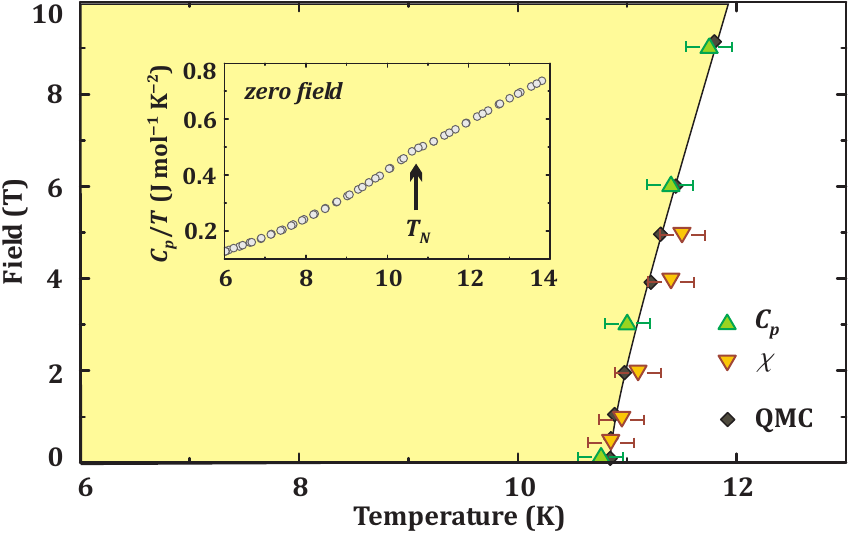}
\caption{\label{fig:diagram}
(Color online) Field-temperature phase diagram derived from QMC simulations for weakly coupled square planes with an effective interlayer coupling $J_{\perp}^{\eff}/J_1=6.5\times 10^{-3}$. Experimental values of $T_N$ are shown according to the magnetic susceptibility ($\chi$) and specific heat ($C_p$) data. Left (colored) and right (white) parts are the magnetically ordered and paramagnetic phases, respectively. The inset shows the specific heat measured in zero field.
}
\end{figure}

The shape of the susceptibility curve below $T^{\max}$ is a strong evidence for the formation of vacancies in the magnetic layers of diaboleite. Although 3D correlations (interplane couplings) might have similar effect on the susceptibility, their contribution is not strong enough to explain the experimental data. For example, weakly coupled square planes with an effective interlayer coupling $J_{\perp}^{\eff}/J_1=6.5\times 10^{-3}$, which reproduces the magnetic transition temperature in diaboleite (see Fig.~\ref{fig:diagram}), show nearly the same susceptibility as the single square plane. 

The antiferromagnetic ordering in diaboleite manifests itself by a kink of the magnetic susceptibility (Fig.~\ref{fig:chi-field}). The N\'eel temperature increases with the applied field. The magnetic transition is not well seen in zero-field specific heat, although a close examination of the $C_p/T$ data at $11-12$~K spots a tiny feature at 10.7~K (see the inset of Fig.~\ref{fig:diagram}). Otherwise, the temperature dependence of the zero-field specific heat is smooth up to at least 100~K.\cite{supplement} A magnetic field enhances the transition anomaly that becomes well visible at 6~T and 9~T (see the inset of Fig.~\ref{fig:chi-field}).

The broadening of the transition anomalies, which is inevitable in powder samples, prevents us from the precise evaluation of $T_N$ in different magnetic fields. Nevertheless, the data in Fig.~\ref{fig:chi-field} show clearly that $T_N$ increases with the applied field, and this trend holds up to at least 9~T. Using middle points of the local minima of the susceptibility, as well as the kinks in the specific heat data, we tentatively evaluate the field dependence of $T_N$ (Fig.~\ref{fig:diagram}). The steady increase in $T_N$ is very typical for spin-$\frac12$ quantum magnets and resembles the field-induced behavior observed in other square-lattice systems with spin-$\frac12$.\cite{tsirlin2011,sengupta2009}

To analyze the field dependence of $T_N$, we first consider the magnetic transition in zero field. The 3D AFM order is driven by the interlayer couplings $J_{\perp}$ and $J_{\perp}'$. These couplings are comparable in strength (see Table~\ref{tab:couplings}) and frustrated, because they favor different interlayer spin configurations: G-type (antiparallel spins on nearest neighbors along $c$) and C-type (parallel spins on nearest neighbors along $c$), respectively. However, the DFT evaluation of $J_{\perp}$ and $J_{\perp}'$ comes with a caveat that the couplings on the order of 1~K are very difficult to estimate precisely, and a careful experimental work, such as inelastic neutron scattering at very low energies, would be necessary to get accurate values of $J_{\perp}$ and $J_{\perp}'$. Here, we argue that the experimental data are in agreement with an effective non-frustrated interlayer exchange that might be a combination of $J_{\perp}$ and $J_{\perp}'$.

We have studied the AFM ordering in weakly coupled square planes by QMC simulations of a 3D spin model with the uniform interlayer coupling $J_{\perp}^{\eff}$ (see Refs.~\onlinecite{sengupta2009} and~\onlinecite{yasuda2005} for similar studies). The $J_{\perp}^{\eff}$ parameter was adjusted as to match $T_N\simeq 10.7$~K in zero field. The resulting estimate of $J_{\perp}^{\eff}/J_1=6.5\times 10^{-3}$ (i.e., $J_{\perp}^{\eff}\simeq 0.2$~K) is somewhat smaller than $J_{\perp}$ and $J_{\perp}'$ (Table~\ref{tab:couplings}) and may reflect their partial compensation. Using this value of $J_{\perp}^{\eff}$, we are able to reproduce both $T_N$ in zero field and the field dependence of $T_N$ (Fig.~\ref{fig:diagram}). This way, the experimental data available so far are in agreement with the spin model of weakly coupled square planes without any substantial frustration. Regarding the role of Cu vacancies and stacking faults, their effect on the interlayer exchange is likely minor and can not be distinguished using the experimental data available so far.

The physical mechanism behind the increase in $T_N$ is the suppression of quantum fluctuations by the applied magnetic field. In higher fields, this effect is countered by the increasing tendency toward parallel spin arrangement (see, e.g., Ref.~\onlinecite{sengupta2009}). However, the AFM couplings in diaboleite are strong enough to compete with the field, so that the $T_N$ increases linearly up to at least 9~T. 

\section{Discussion and Summary}
\label{sec:discussion}
Our study of diaboleite raises several problems that may have implications for other quantum magnets and for theoretical models of quantum magnetism. First, our anticipation of the sizable second-neighbor coupling $J_2$ was not confirmed in the experimental and computational study. While diaboleite indeed bears structural similarity to PbVO$_3$, which is the strongly frustrated square-lattice compound with sizable $J_2$,\cite{tsirlin2008} details of their electronic structures are notably different. In PbVO$_3$, vanadium polyhedra are directly connected to each other, thus leading to relatively short pathways for $J_1$ (V--V distance of 3.88~\r A) and $J_2$ (5.49~\r A). The diaboleite structure features isolated CuO$_4$ plaquettes, hence both $J_1$ and $J_2$ correspond to much longer Cu--Cu distances of 5.88~\r A and 8.32~\r A, respectively. 

The substantial elongation of the metal--metal distances in diaboleite could be mitigated by a different symmetry of the magnetic $3d$ orbital. In contrast to the half-filled $d_{xy}$ orbital of V$^{+4}$, the $d_{x^2-y^2}$ orbital of Cu$^{+2}$ features the strong $\sigma$-overlap with the oxygen $p$ orbitals. This way, the Wannier functions are extended by large oxygen contributions (``tails'') that can overlap and eventually lead to strong couplings even for very long superexchange pathways. In diaboleite, this mechanism is indeed operative for $J_1$, where the O--O distance of 3.13~\r A results in a sizable coupling of 35~K despite the $90^{\circ}$ overlap, which is generally deemed unfavorable for the AFM coupling. Regarding $J_2$, the O--O distance of 4.43~\r A appears to be too long for any appreciable magnetic interaction, even though the orbitals form the favorable $180^{\circ}$ geometry (Fig.~\ref{fig:wannier}). 

The second interesting aspect of the diaboleite magnetism pertains to experimental signatures of the magnetic transition. While the susceptibility measurements unequivocally demonstrate the 3D magnetic ordering below $T_N\simeq 11$~K (Fig.~\ref{fig:chi-field}), heat capacity measured in zero magnetic field shows only a tiny feature that, when examined on its own, may be well considered an artifact (see the inset of Fig.~\ref{fig:diagram}). The lack of a conspicuous transition anomaly is caused by the small amount of the magnetic entropy available at $T_N$. Using QMC simulations, we estimate that in zero field only 7\,\% of the magnetic entropy can be released at $T_N$. The main release of the entropy is observed at $T\simeq J$ as a broad maximum in the temperature dependence of the magnetic specific heat.\cite{sengupta2003,makivic1991} An external magnetic field shifts the entropy from the broad maximum at $T\simeq J$ toward the transition anomaly at $T_N$.\cite{nath2008,*tsirlin2011} This way, the transition anomaly becomes visible and can be tracked with heat-capacity measurements. 

Similar effects of low magnetic entropy at $T_N$ and its dramatic increase in the applied field have been reported for the square-lattice magnets with weak exchange couplings and relatively low saturation fields.\cite{nath2008,tsirlin2011,tsyrulin2010} Although diaboleite features the stronger exchange and the high saturation field (expected at $H_s=4J_1g\mu_B\simeq 120$~T), even the low fields of $6-9$~T ($H/H_s=0.05-0.075$) are sufficient to render the transition anomaly visible. Our experimental results are in line with earlier reports on other spin-$\frac12$ materials. For example, Lancaster~\textit{et~al.}\cite{lancaster2007} claimed that quasi-2D systems with weak interlayer exchange do not show transition anomalies in the specific heat, whereas Sengupta~\textit{et al.}\cite{sengupta2003} demonstrated the suppression of the transition anomaly depending on the size of the interlayer couplings. Our data emphasize the importance of heat-capacity measurements in high magnetic fields, because an external field enhances the anomaly and facilitates an experimental observation of the magnetic transitions. This result may be relevant to the controversial magnetic transition in PbVO$_3$ as well.\cite{tsirlin2008,oka2008}

Diaboleite is a rare material showing magnetism of the diluted square lattice with spin-$\frac12$. We have proved the formation of Cu vacancies by XRD and chemical analysis, and observed the dilution effect in magnetic susceptibility measurements (Fig.~\ref{fig:chi}). If the concentration of Cu vacancies could be varied (e.g., by crystal growth in Cu-deficient environment), diaboleite becomes an interesting system with a remarkably ``clean'' mechanism of dilution. Presently, most studies of the diluted square lattice focus on Zn- and Mg-doped La$_2$CuO$_4$, where the doping with non-magnetic atoms leads to a frustration of the spin lattice by the second-neighbor coupling $J_2$.\cite{[{For example: }][{}]liu2009,*carretta2011,delannoy2009} In diaboleite, this effect is suppressed, because the Cu--Cu distances are very large and prevent any long-range superexchange interactions (Fig.~\ref{fig:wannier}). The diluted spin-$\frac12$ systems provide a unique opportunity to study the interplay of disorder and quantum fluctuations.\cite{delannoy2009,vajk2002,chernyshev2001,*chernyshev2002,hoglund2003} Therefore, further studies of diaboleite may be insightful.

More generally, an in-depth physics research performed on natural samples is a delicate compromise between the potential complexity of the sample and the tantalizing opportunity to find unexpected structural and physical effects. Natural samples are prone to contamination by foreign phases and elements. A meticulous sample characterization is an essential part of the work that often ends when the sample is found unsuitable, owing to a contamination by magnetic impurities. However, sample imperfections can be advantageous as well. Sometimes they disclose interesting effects that are neither expected nor observed in synthetic samples. We hope that the study of Cu-based minerals will further contribute to the experimental and theoretical work on quantum magnetism. Detailed characterization of the chemical composition, crystal structure, and physical properties should also advance our understanding of intricate natural processes that lead to the formation of these interesting and aesthetically beautiful materials.

In summary, we have studied the crystal structure and magnetism of the diaboleite mineral. The formation of Cu vacancies, which are tolerated by the crystal structure, leads to the dilution of the spin lattice. This effect is probed directly via chemical analysis and XRD, as well as indirectly via magnetic susceptibility measurements that show clear signatures of the dilution. The concentration of vacancies is about 5\%. Consistently, thermodynamic properties of diaboleite are well described by the spin model of the diluted square lattice with the nearest-neighbor exchange $J_1\simeq 37$~K and 5\% of Cu vacancies. Weak interlayer couplings trigger the long-range magnetic order below $T_N\simeq 10.7$~K in zero field.

\acknowledgments
We are grateful to Gudrun Auffermann for her kind help with the chemical analysis of small mineral samples. We also acknowledge the experimental support by Walter Schnelle and Deepa Kasinathan (thermodynamic measurements), Yurii Prots and Horst Borrmann (laboratory XRD), Yves Watier (ID31), and the provision of the ID31 beamtime by ESRF. We would like to thank the Department of Materials Research and Physics of the Salzburg University for providing the high-quality natural sample of diaboleite from their mineralogical collection (inventory number 2536). AT was supported by the Mobilitas program of the ESF. SL acknowledges the funding from the Austrian Fonds zur F\"orderung der wissenschaftlichen Forschung (FWF) via a Schr\"odinger fellowship J3247-N16.

%

\newpage
\begin{widetext}
\begin{center}
{\large
Supplementary information for 
\smallskip

\textbf{Square-lattice magnetism of diaboleite Pb$_2$Cu(OH)$_4$Cl$_2$}}
\medskip

A. A. Tsirlin, O. Janson, S. Lebernegg, and H. Rosner
\end{center}
\medskip

\begin{figure}[!h]
\includegraphics[width=11cm]{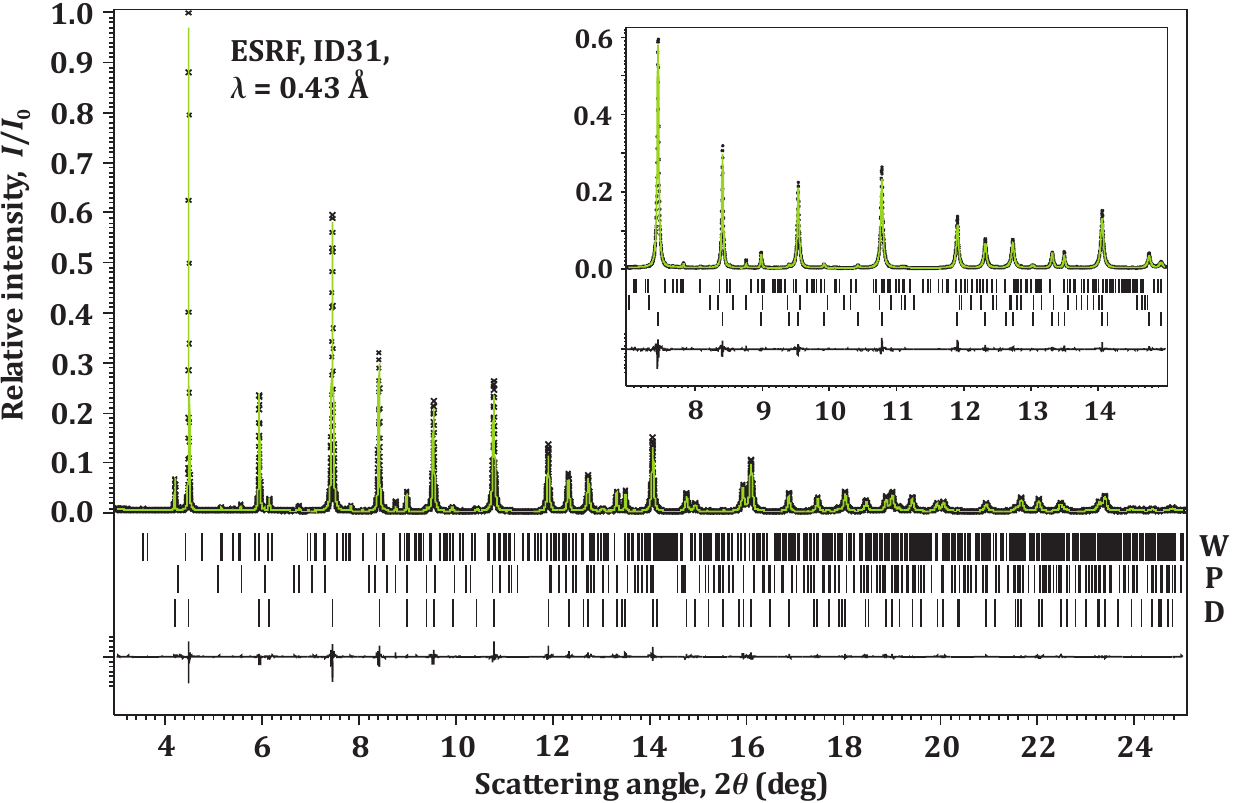}
\begin{minipage}{14cm}
\caption{\label{fig:s1}\normalsize
Rietveld refinement of the room-temperature high-resolution XRD data. Ticks show the reflection positions, from bottom to top: diaboleite (D), phosgenite (P), and wherryite (W). Note that for the diaboleite phase we had to refine the anisotropic strain broadening parameters $S_{hkl}$ (in~10$^{-2}$~deg$^2$~\r A$^{-4}$): $S_{400}=S_{004}=-0.033(2)$, $S_{220}=0.119(3)$, $S_{202}=S_{022}=0.162(3)$, $S_{004}=0$. The definition of $S_{hkl}$ is according to the \texttt{Jana2006} program, see also [P. W. Stephens, J. Appl. Cryst. 32, 281 (1999)].
}
\end{minipage}
\end{figure}

\begin{table}[!hb]
\begin{minipage}{15cm}
\caption{
Atomic positions, atomic displacements parameters $U_{\iso}$ (in 10$^{-2}$~\r A$^2$), and occupation factors $g$ refined from the powder data on the natural sample of diaboleite. Although the accuracy of the structural parameters is probably lower than in [M. A. Cooper and F. C. Hawthorne, Canad. Mineralog. 33, 1125 (1995)], where single-crystal data were used, our refinement provides information on peculiarities of the particular sample used in the present work. Error bars are obtained from the Rietveld refinement.
}
\begin{ruledtabular}
\begin{tabular}{ccccccc}
  Atom & Wyckoff position & $x/a$     & $y/b$     & $z/c$     & $U_{\iso}$ &  $g$      \\
  Pb   &     $2c$         &   0       & $\frac12$ & 0.7298$^a$    &   1.2(1)   &  1.0      \\
  Cu   &     $1a$         &   0       &    0      & 0.008(3)  &   1.0$^b$      &  0.942(4) \\
  Cl1  &     $1a$         &   0       &    0      & 0.464(3)  &   0.1(1)$^c$   &  1.0      \\
  Cl2  &     $1b$	        & $\frac12$ & $\frac12$ & 0.404(2)  &   0.1(1)$^c$   &  1.0      \\
  O    &     $4d$         & 0.2452(6) & 0.2452(6) & 0.9654(8) &   1.3(1)   &  1.0      \\
  H$^d$ &    $4d$         &  0.291    &   0.291   &  0.122    &   1.0      &  1.0      \\
\end{tabular}
\end{ruledtabular}
\begin{flushleft}
$^a$ The $z$ coordinate of the Pb atom is fixed in order to keep the position of origin

$^b$ Only the occupancy factor was refined

$^c$ The atomic displacements parameters of Cl1 and Cl2 were refined as a single parameter

$^d$ The position of hydrogen was not refined
\end{flushleft}
\end{minipage}
\end{table}

\bigskip

\begin{figure}[!h]
\includegraphics[width=9cm]{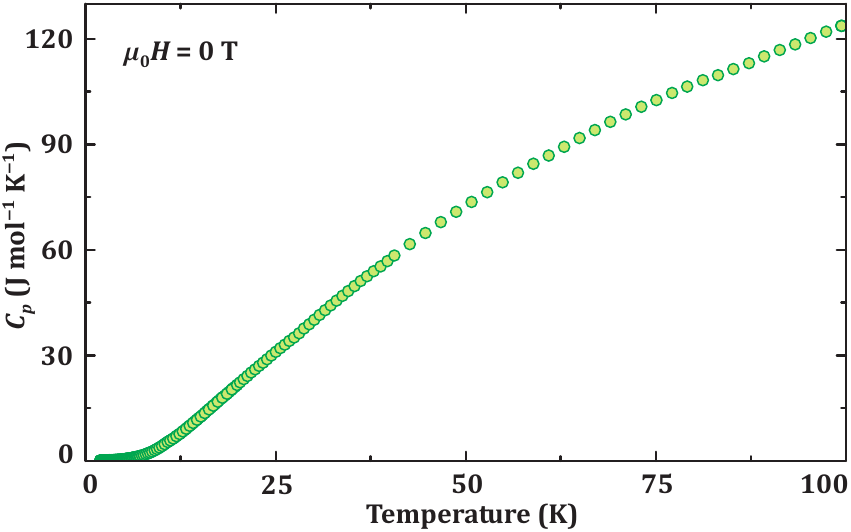}
\begin{minipage}{14cm}
\caption{\label{fig:s2}\normalsize
Zero-field heat capacity of diaboleite.
}
\end{minipage}
\end{figure}

\end{widetext}

\end{document}